\documentclass[aps,prl,reprint]{revtex4-1}

\usepackage{xcolor}
\usepackage{graphicx}
\usepackage{dcolumn}
\usepackage{braket}
\usepackage{bm}

\begin{document}

\preprint{APS/123-QED}

\title{Continuous Raman sideband cooling beyond the Lamb-Dicke Regime in a trapped ion chain}


\author{Qiming Wu}
\email{qiming.wu@nyu.edu}
\author{Yue Shi}
\affiliation{Department of Physics, New York University, New York, NY 10003, USA}
\author{Jiehang Zhang}
\email{jzhang2022@ustc.edu.cn}
\affiliation{1. School of Physical Sciences, University of Science and Technology of China, Hefei 230026, China
}
\affiliation{2. Shanghai Research Center for Quantum Science and CAS Center for Excellence in Quantum Information and Quantum Physics, University of Science and Technology of China, Shanghai 201315, China}
\affiliation{3. Hefei National Laboratory, University of Science and Technology of China, Hefei 230088, China}

\date{\today}

\begin{abstract}
 We report continuous Raman sideband cooling (CRSC) of a long ion chain to the motional ground state beyond the Lamb-Dicke (LD) regime. By driving multiple sideband transitions simultaneously, we show that nearly all axial modes of a 24-ion chain are cooled to the ground state, with an LD parameter as large as $\eta = 1.3$, spanning a frequency bandwidth of 4 MHz. Compared to traditional ground-state cooling methods such as pulsed sideband cooling or electromagnetic-induced-transparency (EIT) cooling, our method offers two key advantages: robustness to timing errors; and an ultra-wide bandwidth unlimited by the number of ions. This technique contributes as a crucial step for large-scale quantum information processing with linear ion chains and higher dimensions alike, and can be readily generalized to other atomic and molecular systems.

\end{abstract}

\maketitle


Trapped atomic ions are a leading platform for quantum computation and simulation~\cite{Debnath2016,pogorelov2021compact,manovitz2022trapped,Zhang2017, jurcevic2014quasiparticle}, precision time keeping~\cite{chen2017sympathetic}, and probing fundamental physics~\cite{sanner2019optical,micke2020coherent}. As a prime candidate for quantum information processing, trapped ions combines high fidelity state preparation and detection~\cite{harty2014high}, long coherence times~\cite{wang2021single}, and high single- and two-qubit gate fidelities~\cite{harty2014high,gaebler2016high,ballance2016high,clark2021high}. These ion qubits can either be coupled in the dispersive regime to realize a quantum simulator with long-range interactions~\cite{porras2004effective,friedenauer2008simulating,kim2010quantum}, or controlled with optimized laser pulses to realize a digital quantum computer~\cite{Debnath2016,pogorelov2021compact,manovitz2022trapped}. In both regimes, quantum entanglement is generated by phonon-mediated interactions, which requires the motional states to be well-defined to avoid incoherent thermal disturbances. Cooling of a trapped ion chain to the motional ground state hence serves as the starting point of high-fidelity quantum logic operations~\cite{gaebler2016high,ballance2016high,clark2021high}.

Resolved sideband cooling is a general method of cooling trapped particles to the quantum harmonic oscillator ground state. It has been demonstrated in various physical systems such as trapped ions~\cite{monroe1995resolved,hamann1998resolved,chan2011laser}, atoms in optical lattices~\cite{vuletic1998degenerate,han20003d}, and  tweezers~\cite{kaufman2012cooling,yu2018motional}. In the case of ions, the conventional setting with tight confinement is referred to as the Lamb-Dicke regime (LDR)~\cite{wineland1998experimental}, in which the motional wavefunction of the ion is well-localized and much smaller than the spatial gradient of the coupling electromagnetic wave. The coupling between the internal electronic and external motional states is weak enough that one can approximate the first-order sidebands as the dominant spin-phonon interaction. Concatenating pulsed red sideband (RSB) excitations and dissipative optical pumping processes~\cite{monroe1995resolved}, ground-state cooling can be achieved after iterating through different mean phonon numbers~\cite{rasmusson2021optimized}. Despite its successful usage in single- and few- ion systems, generalizing this technique to long ion chains meets two crucial challenges: first, sequential cooling of all collective modes of the coupled harmonic oscillator system becomes slow as the system size increases~\cite{chen2020efficient}; second, certain motional modes such as the low-frequency axial modes of a linear chain occupies high phonon numbers, which is detrimental to entangling gates in both axial- and transverse-mode schemes~\cite{pogorelov2021compact,cetina2022control}. Finally, sideband cooling beyond the LDR becomes nontrivial, where the coupling light is strongly modulated by the particle's motion and the cooling efficiency is limited by the large photon-recoil effects~\cite{morigi1997ground}.

In the face of these challenges, various methods have been employed to improve the cooling efficiency and bandwidth: coupling to higher-order sidebands allows faster cooling~\cite{wan2015efficient,yu2018motional}, at the expense of complicated pulse optimizations~\cite{rasmusson2021optimized}; EIT cooling utilizes Fano-like resonances arising from laser-atom interactions to engineer coherent dark-states for suppressing unwanted excitations~\cite{lechner2016electromagnetically,feng2020efficient,qiao2021double,jordan2019near},  with a bandwidth limited by the atomic structure; and polarization gradient cooling serves as a fast intermediate step without cooling to the ground state~\cite{ejtemaee20173d, joshi2020polarization, li2022robust}, thus requiring further sideband cooling. While each method has respective advantages, a single-step technique that combines robustness, simplicity, low temperatures, and high bandwidth is highly desirable. Moreover, a high LD parameter finds many applications, such as enabling faster entangling operations~\cite{sorensen2000entanglement,schafer2018fast,wong2017demonstration}, enhancing quantum sensing~\cite{mccormick2019quantum}, and accelerating novel n-qubit quantum gates~\cite{katz2022n}. Solving the challenge of ground motional state preparation in the presence of a high LD parameter hence becomes an important prerequisite for these new applications. The case of beryllium ions (Be$^{+}$) well illustrates such a scenario: the light mass and clean atomic structure have enabled high-fidelity entangling gates~\cite{gaebler2016high} and large-scale quantum simulations~\cite{britton2012engineered}; but for long linear chains with dozens of Be$^{+}$ ions, the axial modes can span over several MHz of bandwidth with the largest LD parameters exceeding 1, making ground-state cooling for all modes challenging with existing methods. 

\begin{figure}[t]

\includegraphics[width=0.48\textwidth]{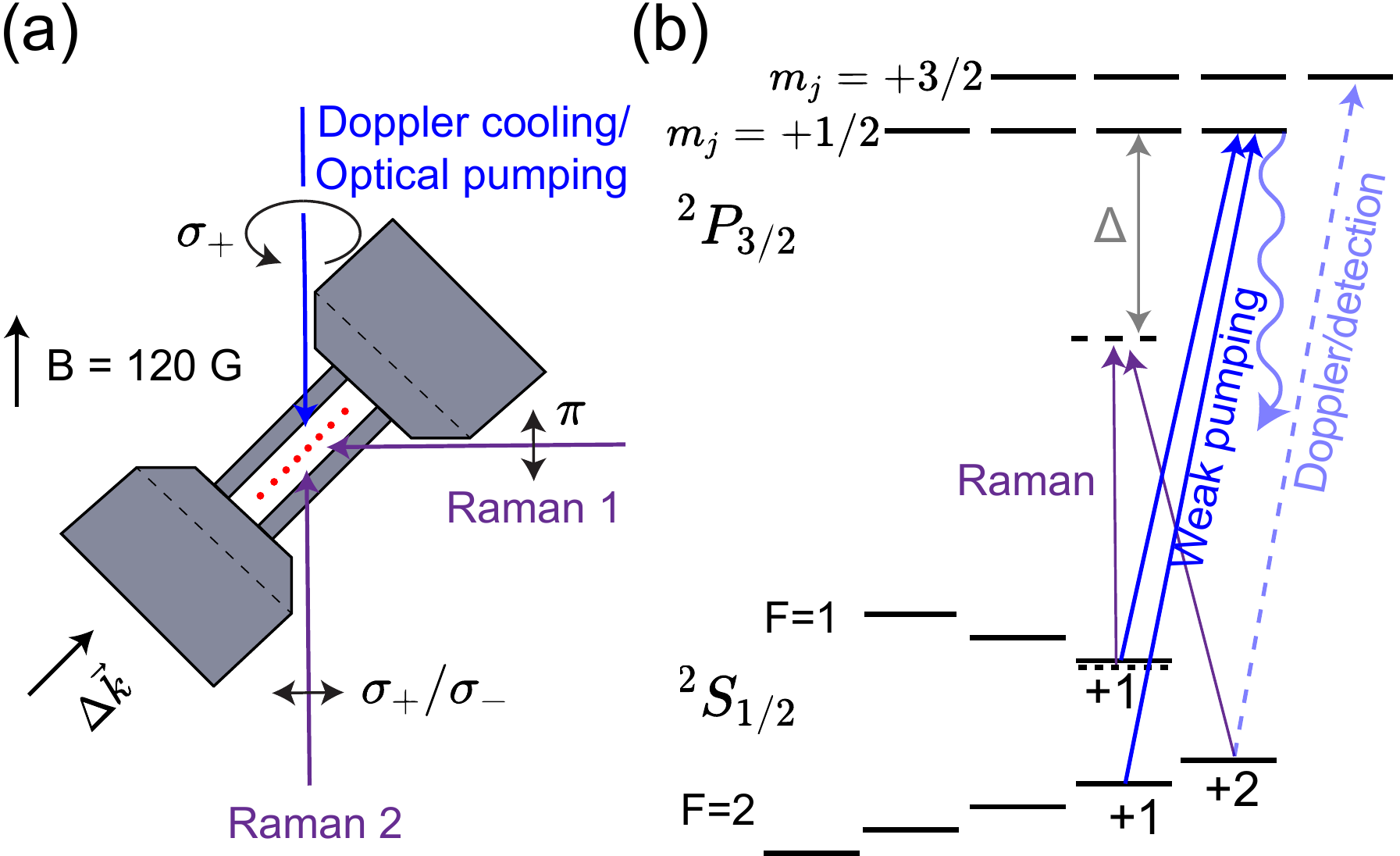}
\caption{Experimental setup and $\mathrm{^9Be^+}$ atomic levels. (a) Ion trap and laser beam configurations: 4-rod trap holds a linear chain under a 120 G magnetic field; a 3-tone Doppler beam with a pure $\mathrm{\sigma_{+} }$ polarization provides Doppler cooling, spin-state detection, and optical pumping. Raman 1 and 2 with $\mathrm{\pi}$ and $\mathrm{\sigma_{\pm} }$ polarizations address the qubit transition with a differential momentum along the trap axial direction. (b) Atomic energy levels involved in the CRSC (not to scale). A pair of multi-tone Raman beams drive RSB transitions in parallel on the $\ket{\downarrow}\leftrightarrow\ket{\uparrow}$ transition, while week pumping beams clear out the populations on the $\ket{\uparrow}$ and $\ket{\mathrm{aux}}$ states through dissipation from the $^2P_{3/2}$ state (wavy line).}
\label{fig:illustration}
\end{figure}

In this article, we solve these challenges by presenting a novel continuous Raman sideband cooling (CRSC) scheme akin to that used in optical qubits~\cite{roos1999quantum}: we continuously drive multiple RSB transitions, and simultaneously apply weak pumping lights to lower phonon occupancy and reset the qubits. Contrasting to the pulsed regime, where a coherent step is followed by a dissipative step, our scheme cools on all phonon states simultaneously and can be thought of as a continuous quantum Zeno process~\cite{itano1990quantum} directing towards the motional ground state. The CRSC is also less susceptible to phonon state distribution and population trapping in certain number states with weak first-order sideband coupling~\cite{yu2018motional}, thus driving higher-order RSBs is not necessary for ground state cooling even far outside the LDR. With this technique, we demonstrate cooling of a single ion to the ground state within 200 $\mathrm{\mu s}$ and efficient cooling of long linear chains containing up to 24 $\mathrm{^9Be^+}$ ions, with LD parameters as large as $\eta = 1.3$, spanning a frequency bandwidth of 4 MHz.


\begin{figure}[!b]
\centering
\includegraphics[width=0.42\textwidth]{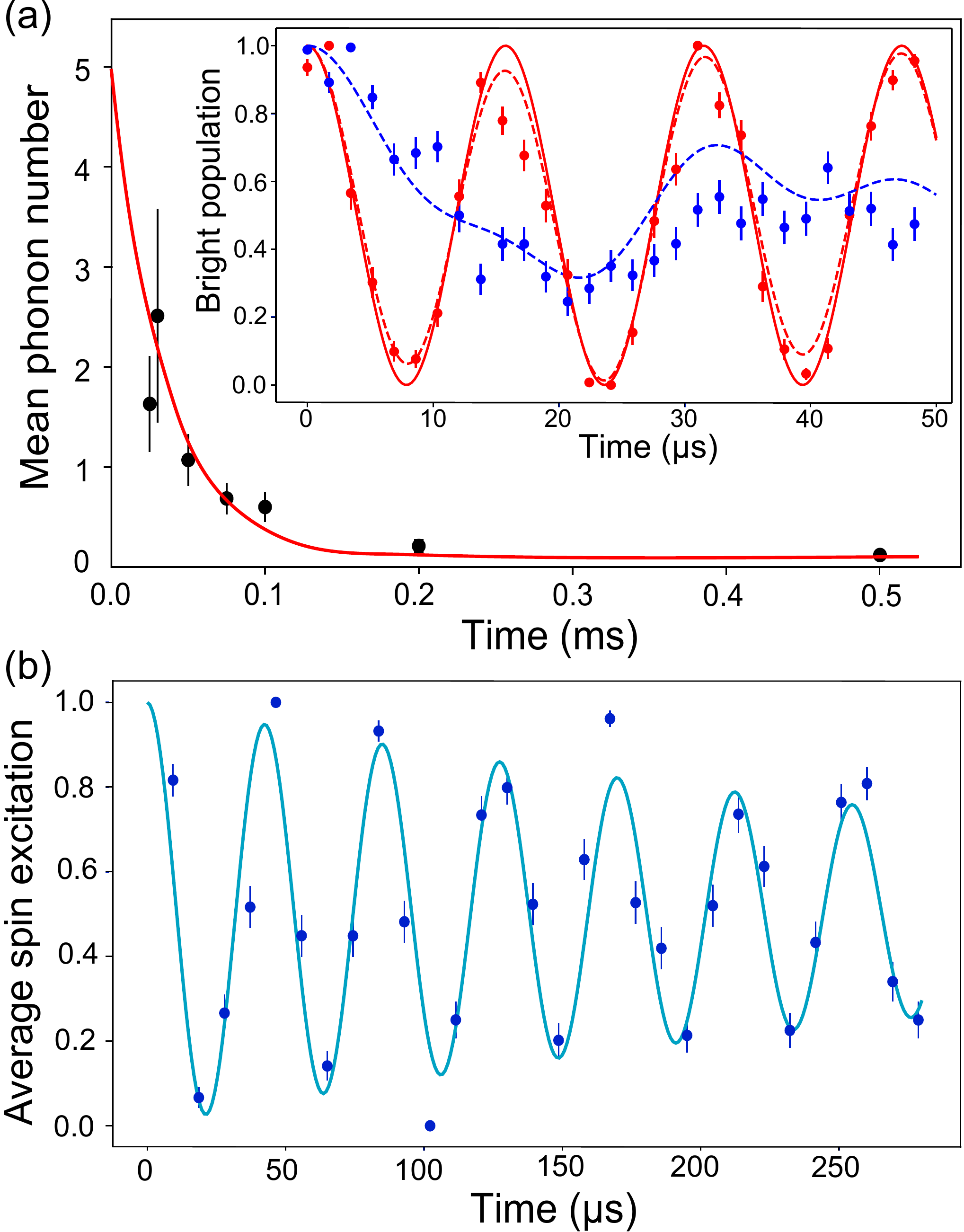}
\caption{Continuous sideband cooling with one and two ions. (a) Cooling dynamics of a single ion. Black points are the measured $\bar{n}$ with the sideband ratio method, error bars denote one standard deviation of the quantum projection noise, and the red line is an exponential fit. Inset shows spin dynamics of carrier transition: red and blue points show the time evolution of spin after CRSC and with only Doppler cooling. The red-solid line shows a sine fit to the data. Red- and blue-dashed lines show the corresponding spin evolution  with $\bar{n}$ = 0.1 and 6, respectively. (b) Average spin excitation of two-ion crystal under the M{\o}lmer–S{\o}rensen interaction. The blue-solid line is a fit of a sine function with an exponential decay envelope.}
\label{fig:single_ion}
\end{figure}

We perform this experiment with ions ablation loaded into a 4-rod radiofrequency (RF) Paul trap~\cite{wu2021adaptively} with a typical radial frequency of $\omega_{x} = 2\pi~\times$ 3.3 MHz and axial frequencies ranging from $\omega_{z} = 2\pi~\times$~270 kHz to $2\pi~\times$~735 kHz. Fig.~\ref{fig:illustration}(a) illustrates the experimental configuration: a $\sigma_+$ polarized Doppler beam near 313 nm is counter-aligned with a 120 G magnetic field defining the quantization axis, with three tones providing Doppler cooling/state detection and two optical pumping frequencies. A pair of global Raman beams with waists of $270\times21~\mu m^2$ propagate at 45 degrees with respect to the trap z-axis, generating a momentum kick $\Delta{\vec{k}}$ along the axial direction. The Raman beam polarizations are tuned to be both linear, one parallel ($\pi$ light) and one perpendicular (equal $\sigma_+/\sigma_-$ light) to the magnetic field, minimizing the vector Stark shift for our Zeeman qubit operations~\cite{wineland2003quantum}. Fig.~\ref{fig:illustration}(b) shows the atomic energy levels of $\mathrm{^9Be^+}$ involved in CRSC. We first apply Doppler cooling for 1~ms with step-wise reduced laser power to approach the Doppler limit ($\bar{n}\approx6\sim17$ depending on the axial frequency), followed by initiating the ions on the $\ket{\downarrow}\equiv~^2 S_{1/2} \ket{F=2,m_F=2}$ state. During CRSC, Raman beams 280 GHz red-detuned from the $^2 P_{3/2}$ excited state drive the $\ket{\downarrow}\leftrightarrow\ket{\uparrow}\equiv~^2 S_{1/2} \ket{F=1,m_F=1}$ transition, with the difference frequencies tuned to be on resonant with the red-sidebands at a maximum two-photon carrier Rabi frequency of $\Omega_0 = 2\pi\times300$~kHz. We use an arbitrary waveform generator (AWG) for the RF signal modulating the laser frequencies, covering the entire motional spectrum with equal amplitudes on all other modes except twice the power on the center-of-mass (COM) mode. Meanwhile, we apply week optical pumping beams to clear out the populations on the $\ket{\uparrow}$ and $\ket{\mathrm{aux}}\equiv~^2 S_{1/2}\ket{F=1,m_F=1}$ states via the $\ket{e}\equiv~^2 P_{3/2} \ket{m_I=3/2,m_j=1/2}$ state, with Rabi frequencies of  $\Omega_1 = 2\pi\times700$~kHz and $\Omega_2 = 2\pi\times600$~kHz, respectively (corresponding to saturation parameters s = 0.23 and 0.45). This pumping scheme eliminates the need for a D1 line laser, which further simplifies the beryllium quantum processor setup, requiring only one frequency-stabilized laser. After CRSC, we probe the RSB/BSB transitions with coherent Raman operations, followed by state detection using spin-dependent fluorescence.

We first show the ground-state cooling of a single ion with $\eta=0.78$. After applying CRSC by  driving the first-order RSB and optical pumping transitions, we measure the mean phonon number $\bar{n}$ with the ratio of red- to blue-sideband (BSB) amplitudes $R={P_{\mathrm{rsb}}}/{P_{\mathrm{bsb}}}={\bar{n}}/({\bar{n}+1})$~\cite{leibfried2003quantum} at different sideband cooling durations $\tau_c$ (Fig.~\ref{fig:single_ion}(a)). An exponential fit shows a $1/e$ cooling time-constant of 32 $\mu s$, with a steady-state $\bar{n}$ = 0.10(4) limited by the single photon recoil from optical pumping. Alternatively, we analyze the $\bar{n}$ with the carrier Rabi flopping of a single ion, which is strongly modulated even at low phonon occupations due to the large LD parameter. The spin evolution under a thermal state assumption is written as:
\begin{equation}
    P_e(t)=\frac{1}{2}\left[1+\sum_{n=0}^{\infty}\frac{\bar{n}^n}{(\bar{n}+1)^{n+1}}\cos(\Omega_{n,n}t)\right],
\end{equation}
 where $\Omega_{n,n}$ corresponds to the Rabi frequency of the carrier transition on $\ket{n}$ state. 
 Inset in Fig.~\ref{fig:single_ion}(a) shows the carrier Rabi flopping of a single ion before and after CRSC: the spin coherence of a Doppler-cooled ion quickly disappears because of the thermal modulation, and the spin dynamics after ground-state cooling show coherent sinusoidal oscillations. While in pulsed sideband cooling, the thermal distribution assumption might be inaccurate~\cite{chen2017sympathetic,rasmusson2021optimized}, in our scheme, the two methods for $\bar{n}$ analysis agree well despite the different sensitivities for the detailed distribution. We compare the experimental cooling dynamics with a numerical simulation using the master-equation and find a good agreement of cooling to the ground state in 200 $\mu s$~\cite{wu2022supplemental}.

For cooling a pair of ions, we apply dual RSBs on the COM and stretch modes, serving as the first step towards generalizing to larger systems. We then drive entangling interactions between the two qubits off-resonantly using the bi-chromatic M{\o}lmer–S{\o}rensen scheme~\cite{sorensen2000entanglement} and observe coherent dynamics with a $1/e$ decay time of 390~$\mathrm{\mu s}$ (Fig.~\ref{fig:single_ion}(b)), exceeding the dominant dephasing from single Zeeman qubit $T_2^*$ time of 300 $\mu s$ (measured with a Ramsey sequence), manifesting dynamical decoupling effects due to the collective drive. To the best of our knowledge, this two-ion interaction is under the largest LD parameter reported so far, even exceeding that in ultrafast gates~\cite{schafer2018fast,wong2017demonstration} .


\begin{figure}[!b]

\includegraphics[width = 0.45\textwidth]{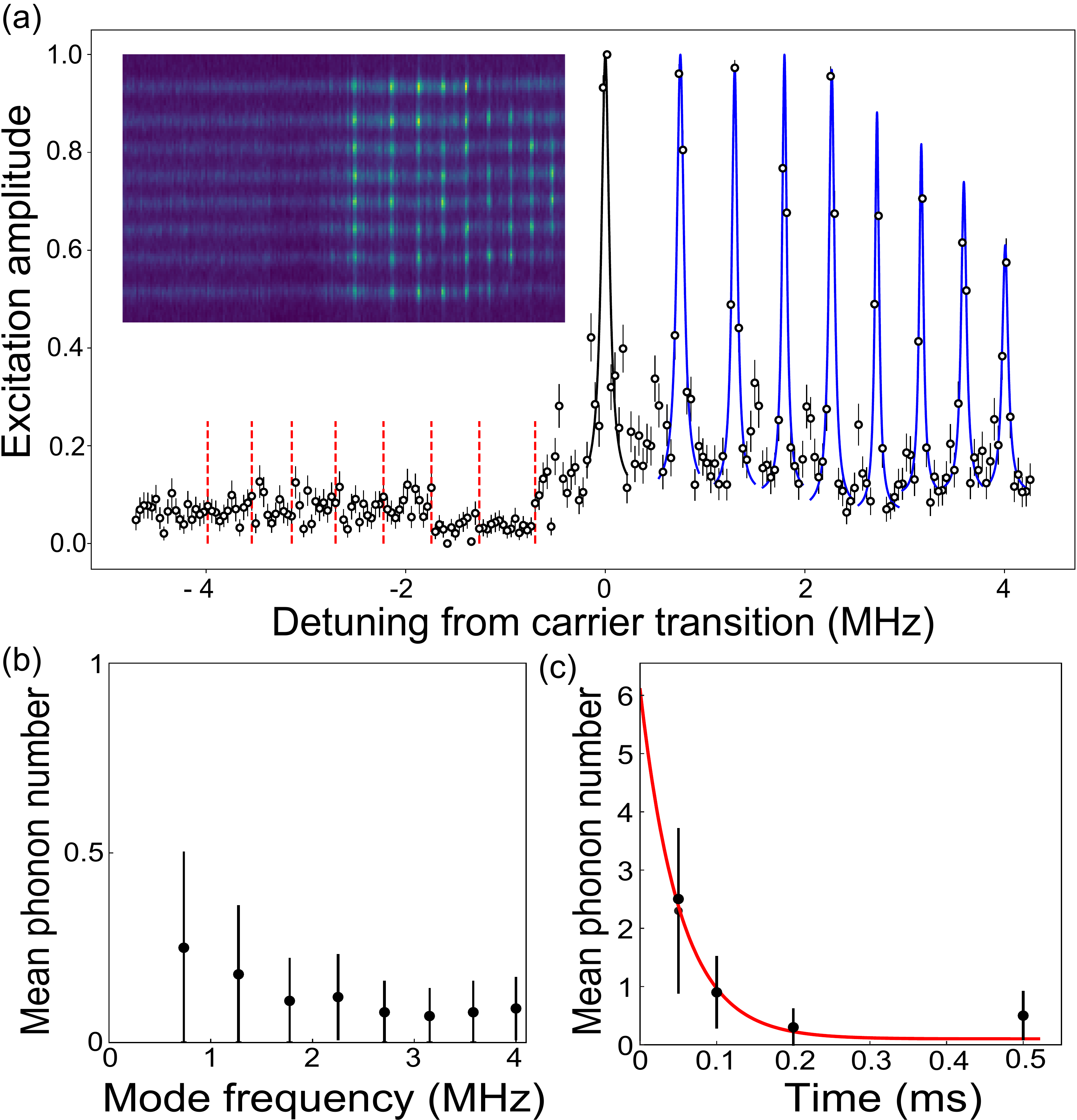}
\caption{CRSC of an 8-ion chain. (a) Motional sideband spectrum of an 8-ion chain after CRSC. The black points are experimental data, and black- and blue-solid lines are Lorentzian fits for the carrier and BSB transitions. Red-dashed vertical lines mark the calculated RSB transition frequencies. Inset shows the frame-averaged ion fluorescence on an EMCCD camera. (b) Extracted mean phonon numbers of the motional modes from the re-scaled fittings of the red-to-blue sideband ratios (see text). (c) Cooling dynamics of the COM mode in an 8-ion chain. Black points are the extracted $\bar{n}$, and the red line is an exponential fit.}
\label{fig:8_ion}
\end{figure}

\begin{figure*}[!t]
\centering
\includegraphics[width=0.92\textwidth]{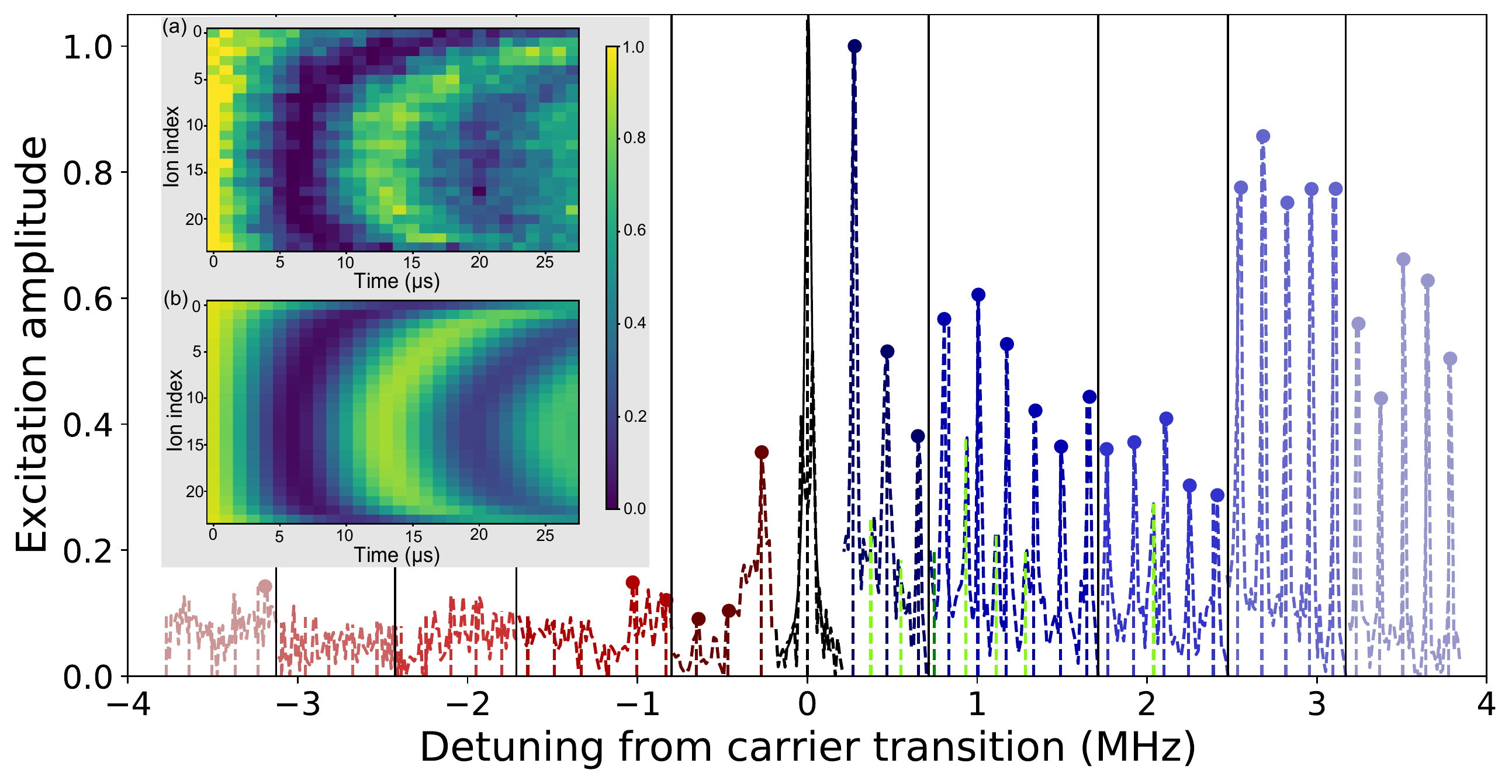}
\caption{The frequency spectrum of a 24-ion chain with COM mode $\eta$= 1.3 after 10 ms of CRSC. Dotted lines are the experimental results of normalized excitation level versus frequency, where red, black, and blue colors denote the scan over RSB, carrier, and BSB transitions, respectively. The brightness of the dotted line shows the variable probe times (30 $\mu s$, 45 $\mu s$, 50 $\mu s$, 60 $\mu s$, 70 $\mu s$ from the darkest to the lightest colors, with black-solid lines separating them) throughout the spectrum to ensure that all mode is excited with high visibilities, compensating the large differences of LD parameters between the low- and high-frequency modes. 
Blue- and red-dashed vertical lines show all the 24 calculated first-order BSB and RSB frequencies. Green-dashed vertical lines show the mode-mixings between COM and a few higher frequency modes, coinciding with the smaller peaks in the experimental data. Inset: (a) spin dynamics of the carrier Rabi flopping after 10 ms of CRSC. The horizontal axis is the probe time, and the vertical axis is the ion index; (b) numerical simulation with a COM mode $\bar{n}$= 1.5, with experimental Raman beam profiles. Note that the carrier Rabi frequency has a strong modulation across the ion chain due to the different coupling factors $C_i = \prod_{m=1}^M e^{-\eta_{i, m}^{2} / 2}$.}
\label{fig:24_ion}
\end{figure*}

We next study the cooling of an 8-ion chain. We program the AWG with eight tones to match the RSB frequencies of the axial vibrational modes. Fig.~\ref{fig:8_ion}(a) shows the motional spectrum of 8 ions after 5~ms of CSRC with a fixed probe time of 15~$\mu s$, which is chosen to excite all modes with different LD parameters with appreciable amplitudes. The data shows near-zero RSB and strong BSB excitations. Inset in Fig.~\ref{fig:8_ion}(a) shows the spin-phonon dynamics on different BSBs imaged on an EMCCD camera. For extracting the $\bar{n}$ of each mode, although the sideband ratio method is valid with individual addressing beams~\cite{chen2020efficient}, it underestimates when the ions are driven with global excitations as in our case ~\cite{lechner2016electromagnetically,qiao2021double}. We sequentially scan over each BSB and corresponding RSB with a probe time that maximizes the BSB transition excitation amplitude, measures the red-to-blue ratio $R_m$, and introduce a scaling factor $\alpha_m$~\cite{lechner2016electromagnetically,qiao2021double}. We then numerically search the best fitted $\bar{n}_m$ of the value $\bar{n}_m = \alpha_m R_m/(1- \alpha_m R_m)$ based on the time evolution of the Hamiltonian of multi-spin coupled to one phonon mode of interest:
\begin{eqnarray}
{\hat{H}_{\mathrm{int}}(t)}=&&\sum_i\frac{\hbar}{2}\Omega_{0} \sigma_{+}^{(i)} \exp \left\{i \eta_{i,m}\left(\hat{a_m} e^{-i \nu_m t}+\hat{a_m}^{\dagger}
e^{i \nu_m t}\right)\right\}\nonumber\\
&&e^{i(-\delta t)} 
+\text { H.c. },
\label{eqn:spin_phonon}
\end{eqnarray}

where $\sigma_{+}^{(i)}$ is the spin-flip operator on ion i, $\eta_{i,m}$ is the LD parameter of (ion i, mode m), $\nu_m$ is the frequency of mode m, and $\delta$ is the laser detuning. We extract
$\bar{n}_m$ for all modes to be below 0.25 (Fig.~\ref{fig:8_ion}(b)), with errors limited by statistical uncertainties. In addition, we measure the cooling dynamics of the COM mode in an 8-ion chain and find the cooling time-constant to be 52 $\mu s$ (Fig.~\ref{fig:8_ion}(c)). Given the same total Raman laser power used for cooling, the experiment rate scale faster than numerical simulations~\cite{wu2022supplemental}($\mathcal{T}\sim O(\ln{n})$), possibly due to additional off-resonant coupling to higher-order modes for multi-ion cases.  


Finally, we extend the CRSC technique to the near-opposite limit of LDR: $\eta > 1$~\cite{morigi1997ground}. For small systems such as one or two ions, sideband cooling through the first-order RSB transition becomes inefficient~\cite{yu2018motional}, and the dynamics are complicated because of the quasi-continuous energy spectrum from mode mixings~\cite{morigi1999laser,jordan2019near}. We apply 5 ms of CRSC on the  the second-order RSB to reach $\bar{n} = 0.27(9)$ for a single ion at $\eta =1.3$, consistent with the recoil limit~\cite{wu2022supplemental}. However, with dozens of ions in a long chain, the first-order modes become dominant again due to the large effective mass, reducing the LD factor for each mode. By addressing all the first-order RSBs and applying CRSC for an extended duration of  10 ms, we demonstrate near ground-state cooling in a 24-ion chain by scanning the sideband spectrum (Fig.~\ref{fig:24_ion}). We observe all 24 first-order BSBs, where the calculated mode frequencies agree well with the experiment. In addition, several small mixing modes between the COM mode and higher frequency modes are visible. 
The RSB transitions are highly suppressed, with near-zero excitations except for the COM mode. Owing to spatial correlations of the electrical field noise, the heating rate for the COM mode scales with the ion number linearly, with equilibrium $\bar{n}$ balanced between the heating and cooling rates~\cite{joshi2020polarization}. 
Since numerically calculating the scaling factors in the sideband ratio method becomes difficult for such a large quantum system~\cite{mei2022experimental}, we evaluate the cooling effectiveness by analyzing the carrier Rabi flopping. 
Fig.~\ref{fig:24_ion} inset shows the experiment result of 24 spin excitations and the numerical simulation of the spin dynamics with a conservatively estimated COM mode $\bar{n}$ of 1.5, where all other modes are presumed to have ground state occupation. The spatial inhomogeneity partially arises from the Gaussian laser beam intensity distribution, which is taken into account in our simulation, but also reflects the spatial dependence in coupling to different modes at such a high LD parameter: the edge ions have $\sim$16 \% slower Rabi rates than the center ions even at motional ground states, with laser beams homogeneously illuminating the chain.
Our result puts a lower bound on the power of the CRSC technique, where the speed could be further optimized using numerical methods such as machine learning~\cite{vendeiro2022machine}, and the temperature could be lowered by other sub-recoil techniques such as velocity-selective coherent population trapping~\cite{park2022motion}. 



In summary, we demonstrate a continuous Raman sideband cooling scheme scalable for large trapped ion systems on account of its flexibility, robustness, efficiency, and high bandwidth. These advantages over the traditional schemes are extended far outside the Lamb-Dicke regime. Our method can be readily generalized to three dimensional ground state cooling~\cite{monroe1995resolved}, mixed species ion systems~\cite{negnevitsky2018repeated,chen2017sympathetic}, sympathetic cooling of molecular~\cite{wan2015efficient} and highly charged ions~\cite{king2021algorithmic}, and high-dimensional Coulomb crystals~\cite{jordan2019near,joshi2020polarization,qiao2021double}.  It also applies to many other systems such as atoms or molecules in optical tweezer arrays~\cite{kaufman2012cooling,yu2018motional} and optical lattices~\cite{vuletic1998degenerate,han20003d}. Our study presents a crucial step towards large-scale quantum simulation~\cite{Zhang2017} and computation~\cite{cetina2022control} beyond dozens of qubits.

\begin{acknowledgements}
We thank Ye Wang, Wenchao Ge, Yong Wan, and Zijian Ding for helpful discussions and critical reading of the manuscript. 

During the preparation of our manuscript, we become aware of related work on EIT cooling with an improved bandwidth for four ions~\cite{zhang2021parallel}.
\end{acknowledgements}

\bibliography{RSC_reference}
\bibliographystyle{apsrev}

\end{document}


\title{Supplemental material: Multi-mode Ground-state Cooling of a Trapped Ion Chain beyond the Lamb-Dicke Regime}

\author{Qiming Wu}

\email{qiming.wu@nyu.edu}
\author{Yue Shi}
\affiliation{Department of Physics, New York University, New York, NY 10003, USA}

\author{Jiehang Zhang}
\email{jzhang2022@ustc.edu.cn}
\email{jzhang2022@ustc.edu.cn}
\affiliation{1. School of Physical Sciences, University of Science and Technology of China, Hefei 230026, China
}
\affiliation{2. Shanghai Research Center for Quantum Science and CAS Center for Excellence in Quantum Information and Quantum Physics, University of Science and Technology of China, Shanghai 201315, China}
\affiliation{3. Hefei National Laboratory, University of Science and Technology of China, Hefei 230088, China}
\date{\today}

\renewcommand{\theequation}{S\arabic{equation}}
\renewcommand{\thefigure}{S\arabic{figure}}
\renewcommand{\bibnumfmt}[1]{[S#1]}
\renewcommand{\citenumfont}[1]{S#1}

\maketitle

\section{Single ion cooling speed analysis}

In this section, we provide a numerical simulation of cooling speed at various optical pumping power and compare the cooling rate of CRSC with pulsed Raman sideband cooling (PRSC).

\subsection{Cooling speed vs optical pumping power}
\begin{figure}
    \centering
    \includegraphics[width =\textwidth]{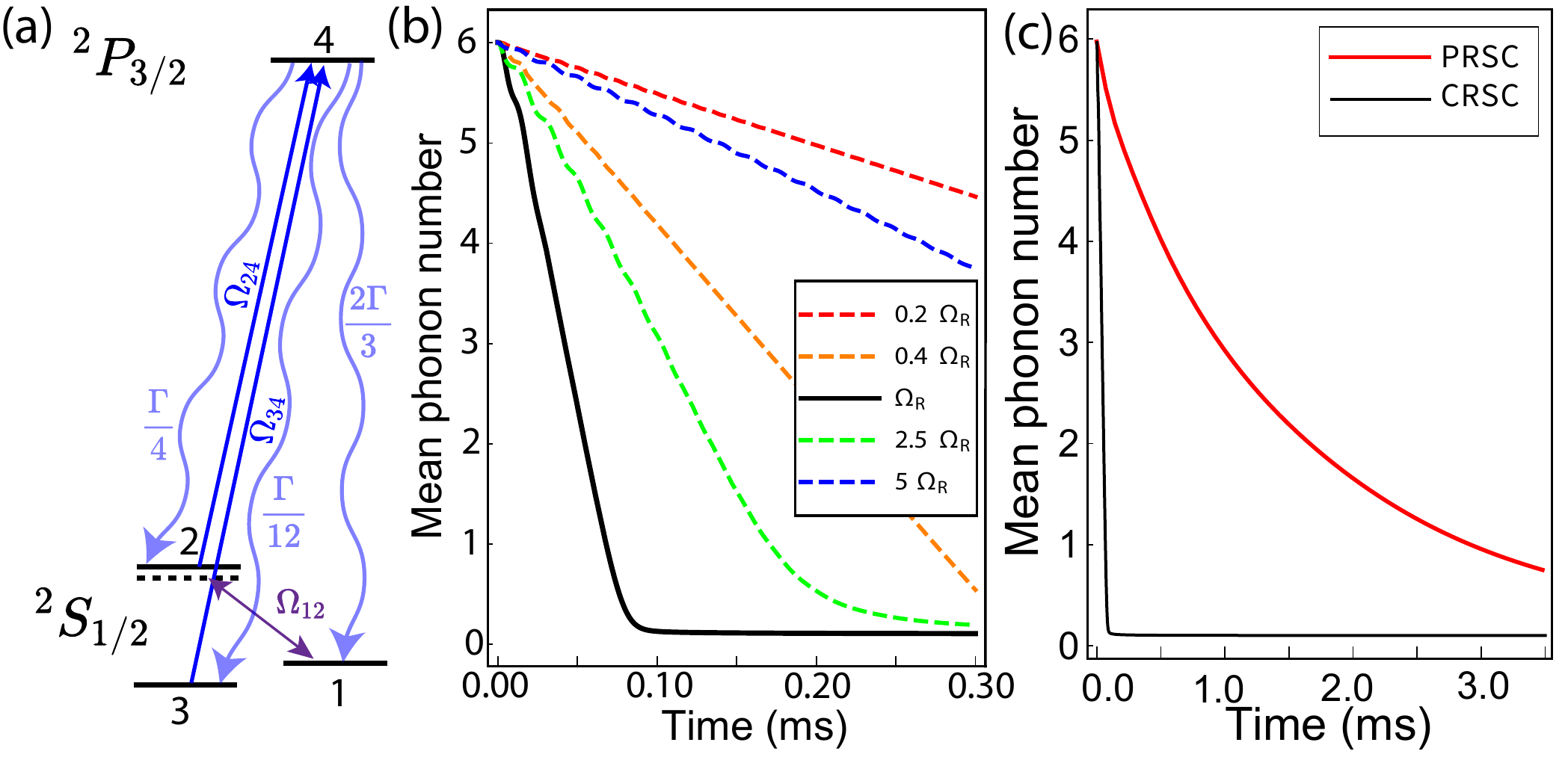}
    \caption{(a) Illustration of the 4-level system for the numerical study of CRSC with the master equation method. (b) Numerical simulation of CSRC dynamics in 300~$\mu s$ of a single ion under different optical pumping Rabi frequencies. (c) Comparison of CSRC dynamics with optimal pumping power and PSRC with optimal pulse length. }
    
    \label{fig:single_ion_cooling_sim}
\end{figure}

To understand the optimal parameters for continuous Raman sideband cooling, we conduct numerical studies and compare them with the experiment. The optical pumping power is critical to the cooling speed and final temperature: it needs to be tuned to a moderate power depending on the carrier Rabi frequency so that the populations quickly reach steady state values. For a single ion, we consider the four relevant atomic levels involved in CRSC to model the cooling dynamics (as shown in Fig.~\ref{fig:single_ion_cooling_sim}(a): $\ket{1}\equiv\ket{\downarrow}$, $\ket{2}\equiv\ket{\uparrow}$, $\ket{3}\equiv\ket{\mathrm{aux}}$ and $\ket{4}\equiv\ket{e}$). Under the RSB Raman drive $\Omega_{12}$ and two optical pumpings $\Omega_{24} $ and $\Omega_{34}$, the Hamiltonian has the form of:
\begin{equation}
    \mathcal{H}=\left(\begin{array}{cccc}
0 & \frac{\Omega_{12}}{2} & 0 & 0 \\
\frac{\Omega_{12}}{2} & 0 & 0 & \frac{\Omega_{24}}{2} \\
0 & 0 & -\delta & \frac{\Omega_{34}}{2} \\
0 & \frac{\Omega_{24}}{2} & \frac{\Omega_{34}}{2} & 0
\end{array}\right).
\end{equation}

Assuming the motion is in a thermal state during the cooling process, $\Omega_{12}$ is the average Rabi frequency of the RSB transition between different number states :
\begin{equation}
    \Omega_{12}(\bar{n}) = \Omega_0 \sum_{n=1}^{\infty}\frac{\bar{n}^n}{(\bar{n}+1)^{n+1}}e^{-\eta^2/2}\eta\frac{1}{\sqrt{n}}L_n^{(1)}(\eta^2),
\end{equation}

where $\bar{n}$ is the mean phonon number, $\eta$ is the LD parameter, and $L_n^{(1)}(\eta^2)$ are the generalized Laguerre polynomial. During our experiment, $\Omega_0 = 2\pi\times$300 kHz is the Rabi frequency of the atomic transition. The pumping speed is set to $\Omega_R$  with the Rabi frequencies of the two pumping beams $R_1:\Omega_{24} = 2\pi\times$700 kHz and $R_2:\Omega_{34} = 2\pi\times$600 kHz ($R_2$). $\delta=-2\pi\times10$~kHz is the detuning of $R_2$ beam from atomic resonance to avoid coherent population trapping. 
To include spontaneous emission from the $\ket{4}$ state, the system evolution can be described by Lindblad equation:
\begin{equation}
    \frac{\mathrm{d} \rho}{\mathrm{d} t}=\frac{1}{i \hbar}[\mathcal{H}, \rho]+\mathcal{K} \rho
     \label{eqn:Lindblad},
\end{equation}

where $\rho$ is the density matrix of the 4-level system, $\mathcal{K} \rho=-\frac{1}{2} \sum_{m=1}^{3}\left[\hat{C}_{m}^{\dagger} \hat{C}_{m} \rho+\rho \hat{C}_{m}^{\dagger} \hat{C}_{m}+\hat{C}_{m} \rho \hat{C}_{m}^{\dagger}\right]$ denotes the dissipation through the three decay channels $\hat{C}_{1}= \sqrt{\Gamma\times 2/3}\ket{1}\bra{4},\hat{C}_{2}= \sqrt{\Gamma/4}\ket{2}\bra{4},\hat{C}_{3}= \sqrt{\Gamma/12}\ket{3}\bra{4} $. $\Gamma$ is the natural linewidth of the $\mathrm{^2 P_{3/2}}$ state of $\mathrm{^9 Be+}$.
The cooling rate approximately equals the number of photons scattering from $\ket{4}$ to $\ket{1}$ state:
\begin{equation}
    \frac{\mathrm{d} \bar{n}}{\mathrm{d} t}\approx- \frac{\Gamma_{41}}{3}\rho_{44}(t)
    \label{eqn:nbar_evolution}.
\end{equation}

We extract the time evolution of $\bar{n}$ from the numerical solves of Eqn.~\ref{eqn:Lindblad} and Eqn.~\ref{eqn:nbar_evolution}. To show the influence of optical pumping on the cooling effect, in Fig.~\ref{fig:single_ion_cooling_sim}(b), we compare the single ion cooling dynamics with different optical pumping speeds: 0.2~$\Omega_R$, 0.4~$\Omega_R$, $\Omega_R$, 2.5~$\Omega_R$ and 5~$\Omega_R$, with $\Omega_R$ being the optimal: cooling to the ground state takes about $200~\mu s$, which agrees with the experiment.

\subsection{Cooling speed: continous vs pulsed}

To show the cooling efficiency of our scheme compared to the more traditional pulsed sideband cooling,  we numerically simulate the cooling speeds of CRSC and PRSC of a single ion. We model the PRSC process with the same carrier Rabi frequency of $\Omega_0 = 2\pi\times300$ kHz, LD parameter $\eta = 0.78$ and initial thermal phonon state $\bar{n}=6$. For simplicity, we ignore the optical pumping time and photon recoil effects and set each pulse to equal length $t$, which shows close to optimal cooling times. The phonon population after the $i$-th pulse can be written as a function of the the population after the previous pulse:
\begin{equation}
\begin{split}
    P_i(0) &=  P_{i-1}(0) + \frac{1}{2}P_{i-1}(1)(1-\cos{\Omega_{0,1}t})\\
 P_i(1) &=  \frac{1}{2}P_{i-1}(1)(1+\cos{\Omega_{0,1}t}) + \frac{1}{2}P_{i-1}(2)(1-\cos{\Omega_{1,2}t})\\
 &\vdots \\
 P_i(n) &=  \frac{1}{2}P_{i-1}(n)(1+\cos{\Omega_{n-1,n}t}) + \frac{1}{2}P_{i-1}(n+1)(1-\cos{\Omega_{n,n+1}t}),
\end{split}    
\end{equation}
where $P_i(n)$ denotes the population on the $\ket{n}$ state in the $i$ step, $\Omega_{n,n+1}$ is the Rabi frequency of $\ket{\uparrow,n+1}\leftrightarrow\ket{\downarrow,n}$ transition. We numerically search the optimal pulse length between 0 and 100 $\mu s$ that yields the lowest $\bar{n}$ after 50 cooling pulses and find T = 71.5~$\mu s$. Fig.~\ref{fig:single_ion_cooling_sim}(c) shows the cooling time comparisons, in which PRSC is significantly slower, and the $\bar{n}$ does not reach the ground state after 3.5 ms. An intuitive reason for the speedup is the following: in the CRSC process, the state population will quickly reach a steady state under an appropriate optical pumping rate. This guarantee the fastest pumping rate towards the motional ground state. CRSC is also more robust than PRSC: it avoids complicated optimization of cooling pulses, driving higher-order sidebands when the ions are outside the LDR, and is not sensitive to timing error.

\begin{figure}
    \centering
    \includegraphics[width = 0.7\textwidth]{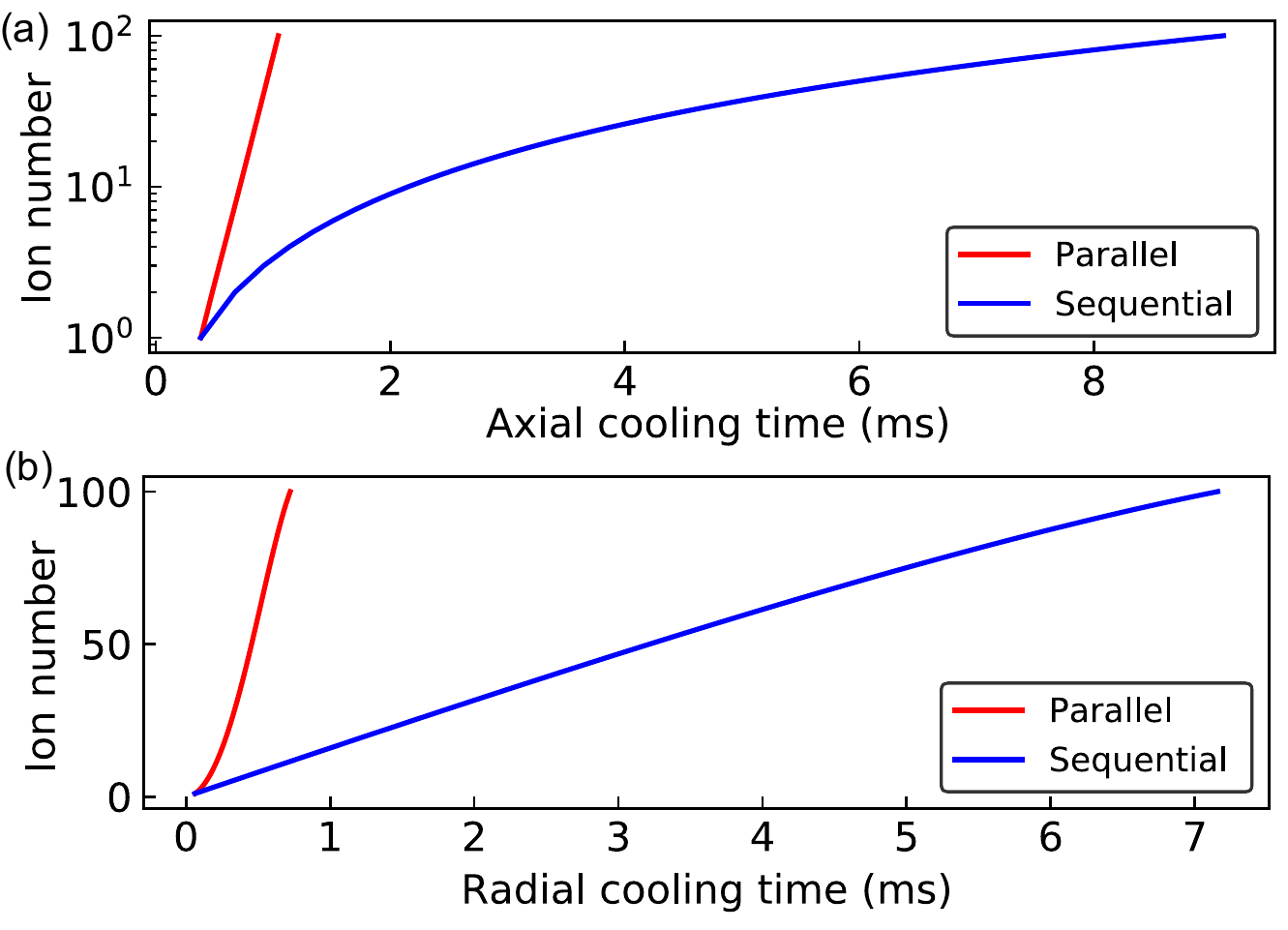}
    \caption{Numerical simulation of (a) axial and (b) radial modes cooling time scaling with the increase of ion number, comparing our parallel with the traditional sequential scheme.}
    
    \label{fig:cooling_speed_discussion}
\end{figure}

\section{Cooling speed scaling with ion number}

Beyond a single ion, we analyze the cooling of multiple modes in a long ion chain and compare the cooling efficiency of our parallel approach to the traditional sequential scheme. The cooling time to the ground state of a single ion is $\mathcal{T}_1 = A\bar{n}_{1}/(\eta\Omega_0) = 200 \mu s$, where A is a constant, $\bar{n}_{1}$ is the initial temperature. Generalizing to an N-ion chain, we assume all modes are resolved in frequency, and each mode is cooled to the Doppler limit. The cooling time for mode $m$ $\mathcal{T}_m = A\bar{n}_m/(\eta_m\Omega_m)$, where $\bar{n}_m\propto 1/\omega_m$ is the initial mean phonon number of the mode, $\eta_m\propto 1/\sqrt{\omega_m}$ is the mode LD parameter and $\Omega_m\propto\sqrt{P_m}$ is the Rabi frequency component on the mode. The total cooling time can be written as : 
\begin{equation}
    \mathcal{T} = \max _{m}\left(\frac{C}{\sqrt{\omega_1 P_1}},\frac{C}{\sqrt{\omega_2 P_2}},\dots,\frac{C}{\sqrt{\omega_N P_N}}\right)
\end{equation}

with $\sum_m P_m = P_0 $. The total cooling time is the shortest $\mathcal{T}_{min}$ when $\omega_1 P_1=\omega_2 P_2=\dots =\omega_N P_N$. So we have $\mathcal{T}_{min}$ cooling axial modes of an N-ion chain:

\begin{equation}
    \mathcal{T}_{min}^N = \sqrt{\omega_1\sum_{m=1}^N\frac{1}{\omega_m}}\mathcal{T}_1\sim O(\ln{N})\mathcal{T}_1,
    \label{eqn:parallel}
\end{equation}

which the cooling speed scaling of 8 ions compared to a single ion agrees with the experiment. In the traditional scheme, the total cooling time $\mathcal{T}_{min}^{\prime N}$ equals the sum of the time spent in cooling each mode:
\begin{equation}
    \mathcal{T}_{min}^{\prime N} = \sqrt{\omega_1}\sum_{m=1}^N\frac{1}{\sqrt{\omega_m}}\mathcal{T}_1\sim O(N)\mathcal{T}_1.
    \label{eqn:sequetial}
\end{equation}
Compared to traditional sequential cooling, our scheme provides an exponential speedup for axial modes in the ion number scaling, which is especially powerful when cooling large ion crystals. To illustrate this, we compare the cooling time with the increase of ion number of parallel versus sequential cooling in our numerics. We assume single ion motional frequency $(\omega_{x},\omega_{z}) = 2\pi\times(7.5, 0.2)$~MHz to maintain a linear chain when trapping many ions. Based on (\ref{eqn:parallel}) and (\ref{eqn:sequetial}), in Fig.~\ref{fig:cooling_speed_discussion} we calculate the cooling time scaling of (a) axial and (b) radial modes up to 100 ions. While our scheme for cooling the axial modes has exponential speedup, cooling for the radial modes scale as $O(N)$, similar to the traditional method. However, the total cooling time is less than 1 ms which makes the scheme scalable. Furthermore, we expect the cooling time to be independent of ion number if all modes are not resolvable.

\begin{figure}[!b]
   \centering
    \includegraphics[width = 1 \textwidth]{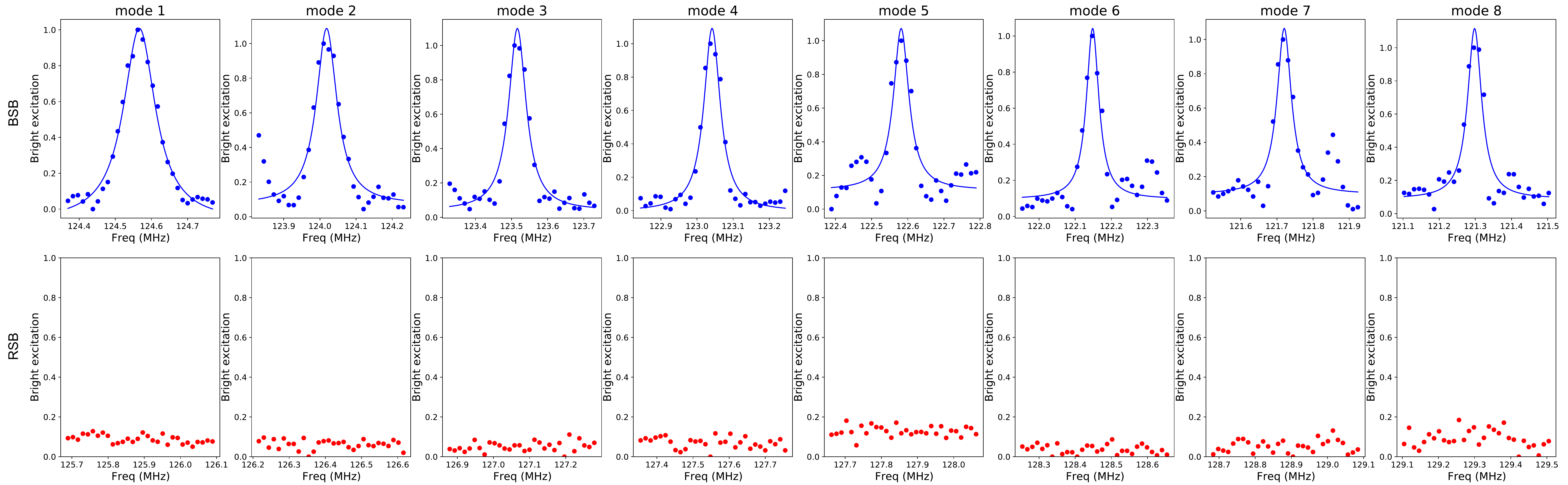}
    \caption{Frequency scan of 8-ion chain on BSB and RSB transitions. Blue and red points are the experimental data for BSB and RSB transitions, respectively. Blue solid lines are Gaussian fits to the BSB data.}
    
    \label{8ions_freq_scan}
\end{figure}

\section{Measurement of multi-ion phonon numbers }
In this section, we discuss two methods to assess the cooling effects of CRSC. 
\subsection{Multi-ion phonon numbers extraction from sideband ratio}
Following the discussion in the main text, the traditional sideband ratio method underestimates the phonon number for the multi-ion case. Therefore, we introduce a scaling factor $\alpha$ to compensate for the underestimation. To further understand the multi-spin phonon dynamics, we numerically study the time evolution of RSB and BSB transitions on each mode, which the Hamiltonian can be described as:

\begin{equation}
\begin{split}
H_{rsb,m} &= (\hbar/2)\Omega_0 \sum_{i=1}^{N}\sigma_{+}^{(i)}(\eta_{i,m}a_m+\eta_{i,m}^3 a_m^2a_m^{\dagger}/2) +\mathrm{H.c.} \\
H_{bsb,m} &= (\hbar/2)\Omega_0 \sum_{i=1}^{N}\sigma_{+}^{(i)}(\eta_{i,m}a_m^{\dagger}+\eta_{i,m}^3 a_m^{\dagger~2} a_m/2) +\mathrm{H.c.}
\end{split}.\label{eq:label2}
\end{equation}

Here, the Pauli operator $\sigma_{+}^{(i)}=1/2(\sigma_{x}^{(i)}+i\sigma_{y}^{(i)})$ acts on the $i$-th ion, $a_m$ and $a_m^{\dagger}$ are the creation and destroy operator on the $m$-th mode, $\eta_{i,m}$ is the Lamb-Dicke parameter of $i$-th ion, $m$-th mode. Here we keep the Lamb-Dicke expansion to the third order. We assume the phonons are thermally populated with mean phonon number $\overline{n_{m}}$. Then we apply a numerical simulation of the time evolution under RSB/BSB drive with the initial state:

\begin{equation}
    \ket{i}_{\overline{n_{m}}} = \ket{\uparrow\uparrow…\uparrow}\otimes\sum_{n_m}\frac{\overline{n_{m}}^{n_m}}{\left(\overline{n_{m}}+1\right)^{n_{m}+1}}\ket{n_m}. 
\end{equation}

The result states  $\ket{r}_{\overline{n_{m}}}(t)$ and $\ket{b}_{\overline{n_{m}}}(t)$. have brightness of $A_{\overline{n_{m}},r}(t)$ and  $A_{\overline{n_{m}},b}(t)$. After T, the BSB flops to the brightest state, and we can calculate the RSB to BSB amplitude:
\begin{equation}
    R(\overline{n_{m}}) = A_{\overline{n_{m}},r}(T)/A_{\overline{n_{m}},b}(T),
\end{equation}

which is a function of the mode index m and the corresponding mean phonon number. In the case of a single ion $R = \overline{n}/(\overline{n}+1)$. We experimentally measure the RSB to BSB amplitude when BSB is the brightest $R_{m,exp}$ and extract $\overline{n_{m}}$ from $R^{-1}(\overline{n_{m}})$. In other words, we introduce a scaling factor $\alpha_m$ so that the mean phonon number of the mode can be calculated as $\bar{n}_m = \alpha_m R_m/(1- \alpha_m R_m)$. $\alpha_m$ is mode-specific and phonon-sensitive ($\alpha=1$ for single ion case). In Fig~\ref{8ions_freq_scan} we show the experimental data of  BSB and BSB floppings. The BSB peaks are normalized to 1 for the excitation amplitude of each mode. RSB peaks are not visible, and we use half of the peak-to-peak noise to give an upper bound of the RSB excitation amplitude $A_{\overline{n_{m}},r}(T)$. From the frequency spectrum, we extract $\bar{n} = \{ 0.25(24),0.18(18),0.11(11),0.12(11),0.09(9),0.07(7),0.08(8),0.09(8)\}$ from the lowest to highest order mode of an 8 ion chain. Although this method has large uncertainty compared to the single ion case, it puts an upper bound on the final temperature without using individual addressing beams.

\subsection{Multi-ion carrier flopping}
Besides scanning the frequency spectrum, analyzing the carrier Rabi flopping is an alternate method to evaluate the cooling effect. Multi-ion carrier flopping under thermal motion can be complicated as ions can be modulated by phonons from all the modes. Now we consider the ion $i$ carrier transition, which couples to all the M modes. In the LD regime a good approximation of the Rabi frequency is $\bar{\Omega}_{i}=\Omega_{i} \exp \left[-\sum_{m} \eta_{i m}^{2}\left(\bar{n}_{m}+1 / 2\right)\right]$. And one can evaluate the cooling effect from the contrast of carrier flopping. However, extracting the mean phonon number from a certain mode is hard. And the dynamics become non-trivial when ions are far outside the LD regime as in our case. The carrier Hamiltonian can be written as:

\begin{equation}
    H_{i,M} = (\hbar/2)\Omega_{i}\sigma_{+}^{(i)}\prod_{m = 1}^M \exp \left[i \eta_{i, m}\left[a_{m}+a_{m}^{+}\right)\right].
\end{equation}

For a specific number state $n_1,n_2,...n_M$, the carrier Rabi frequency equals to:

\begin{equation}    
\begin{array}{lcl}
\Omega^{(i)} & = (\hbar/2)\Omega_{0}^{(i)}<n_1,n_2,…n_M|\prod_{m = 1}^M \exp \left[i \eta_{i, m}\left[a_{m}+a_{m}^{+}\right)\right]|n_1,n_2,…n_M> \\

& = \prod_{m=1}^M e^{-\eta_{i, m}^{2} / 2} \operatorname{L}_{n_m}\left(\eta_{i, m}^2\right)\Omega_{0}^{(i)}.
\end{array}
\end{equation}

\begin{figure}[!t]
   \centering
    \includegraphics[width = 1 \textwidth]{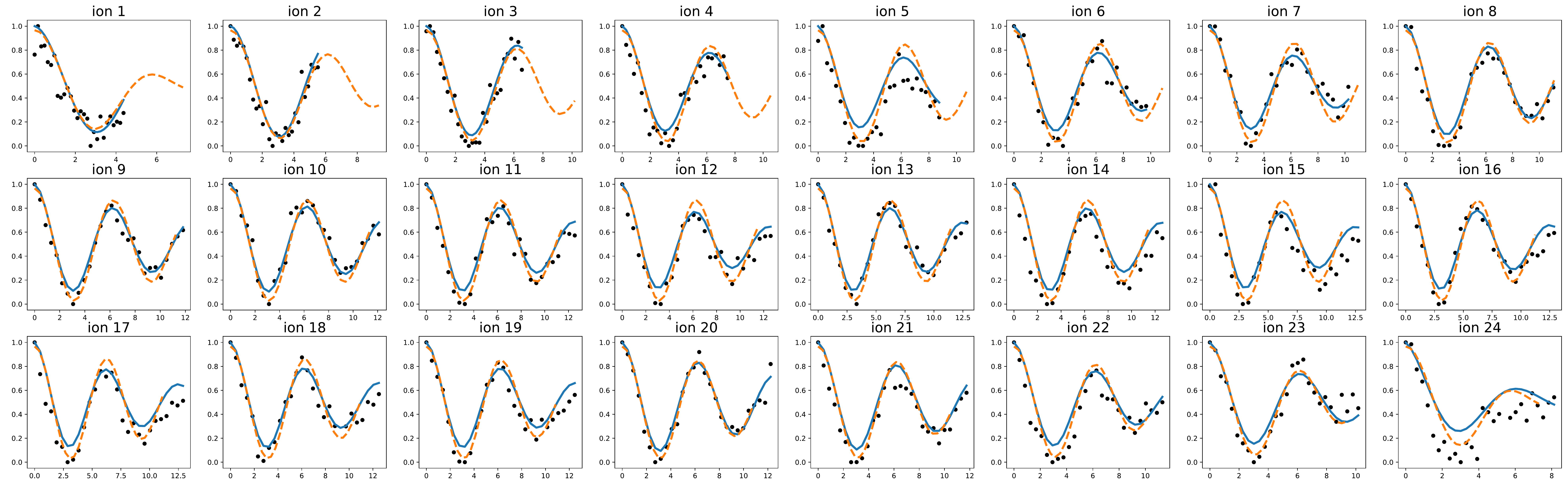}
    \caption{Normalized excitation amplitude of a 24-ion chain on a EMCCD camera. Black points are the experimental data, blue-solid lines are the experimental fit, and orange-dash lines are the numerical simulation of the fitted COM mode $\bar{n} = 1.5$.}
    
    \label{24ions_time_flop}
\end{figure}

If the ions are cooled to the ground state, the carrier Rabi frequency equals to $\prod_{m=1}^M e^{-\eta_{i, m}^{2} / 2}\Omega_{0}^{(i)}$. Note that if the axial COM mode $\eta = 1.3$ of a 24 Be+ chain, the center ions has 19\% stronger Rabi frequency than the edge ions. If we cut off phonon state to n=9, for 24 ions with 24 modes, we need to consider in total $10^{24}$ transitions, each one corresponding to frequency $\Omega_{n,n}$ and phonon state population of:
\begin{equation}
    P_n = \prod_{m=1}^M \frac{\overline{n_{m}}^{n_m}}{\left(\overline{n_{m}}+1\right)^{n_{m}+1}}.
\end{equation}
So that the time evolution of the carrier Rabi flopping equals:
\begin{equation}
    P_i(t) = \frac{1}{2}(1+\sum{(P_n\times\cos({\Omega_{n,n} t)})}.
\end{equation}
From the previous method, since virtually no RSB excitation except the COM mode appears on the 24-ion spectrum, we simplify the model by considering only thermal distribution on the COM mode (discussed in the main text) and assume all other modes are cooled to the ground state. In Fig.~\ref{24ions_time_flop} the black points show the carrier time flop of individual ions. We perform an exponential fit to all the ions, as shown by the blue-solid lines. Then numerically search the best-fitted COM mode $\bar{n}$ to the exponential fits (orange lines).


\section{Cooling limit discussion}
In this section, we discuss several factors which could limit the final cooling temperature.

\subsection{Electric field noise induced heating}
We measure the cooling and heating rate of a single ion's axial motion under the trap frequency $\omega_{1} = 2\pi\times~735$~kHz ($\eta=0.78$) to be $R_{c}^0 =3.2\times 10^4$/s and $R_{h}^0=2.0\times 10^2$/s respectively. The cooling speed $R_c$ scales as $1/\ln{N}$, limited by the laser intensity in each cooling tone and is proportional to the Lamb-Dicke parameter $\eta$. On the other hand, heating of the COM mode scales as $N$ and is proportional to $\omega^{-(\alpha+1)}$. $\alpha$ is the noise intensity at a certain frequency. Considering 1/f noise ($\alpha$=1), we can estimate the cooling and heating rate of the COM mode of a 24-ion chain under the axial frequency of $\omega_2=2\pi\times272$~kHz:
\begin{equation}
\begin{split}
R_{c}^N &= R_{c}^0\sqrt{\omega_{1}/\omega_{2}}/\ln(N) = 1.6\times 10^4 /s\\
R_{h}^N &= NR_{h}^0(\omega_{1}/\omega_{2})^2 = 3.5\times 10^4 /s
\end{split}
\end{equation}

So the equilibrium $\bar{n}\approx R_{h}^N/R_{c}^N = 2.2$, which agrees with the extraction we get from carrier flopping $\bar{n}=1.5$.


    
\begin{figure}[!b]
    \centering
    \includegraphics[width = 0.7\textwidth]{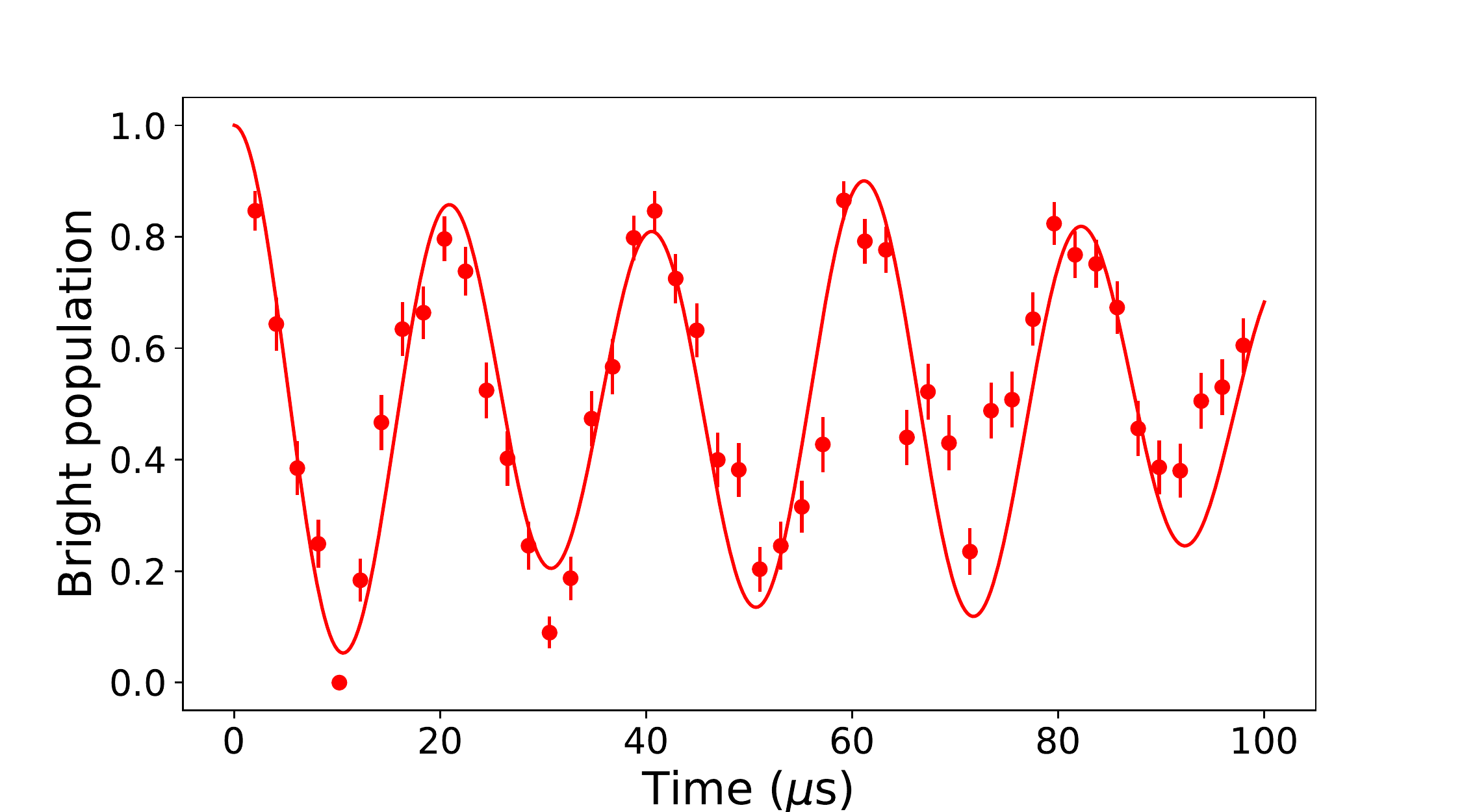}
    \caption{Carrier flopping of single ion at $\eta$ = 1.3 after CSRC. Red points are the measurement data, and the red-solid line is a least-squares fit to the measurements with $\bar{n}=0.27(9)$.}
    
    \label{fig:Single_ion_low_trap}
\end{figure}

\subsection{Photon recoil heating}
Fig.~\ref{fig:Single_ion_low_trap} shows carrier flopping of a single ion at $\eta=1.3$ after 5 ms of CRSC on the second-order RSB,  the measured $\bar{n}=0.27(9)$ is most likely limited by photon recoil effects. We first consider the fundamental cooling limit of photon recoils during CRSC. When the Raman beam drives the qubit on the first-order RSB of $\ket{\downarrow}\leftrightarrow\ket{\downarrow}$ and removes a phonon, optical pumping is required to initiate the qubit in $\ket{\downarrow}$ state, which absorbs and emits 1.5 photons on average. So the heating from spin-reset optical pumping equals:
\begin{equation}
    \begin{aligned}
\Delta{n}_{h}&=(\Delta{E}_{\mathrm{ab}}+\Delta{E}_{\mathrm{em}})/\hbar\omega_z \\
&\simeq1.5\times(\frac{1}{2}+\frac{1}{3}) \tilde{\eta}^2\\
&=1.25\tilde{\eta}^2\,
\end{aligned}
\end{equation}
where $\Delta{E}_{\mathrm{abs}}$ and $\Delta{E}_{\mathrm{em}}$ are the single photon recoil energy of absorption and emission into the axial direction, $\tilde{\eta} = \eta/\sqrt{2}$ is the single photon Lamb-Dicke parameter. So that $\Delta{n}_{h}^{(1)} = 0.38~(\eta=0.78)$ and  $\Delta{n}_{h}^{(2)} =1.1~(\eta=1.3)$. This justifies that at $\eta=0.78$ ground state cooling can be done by driving the first-order RSB, while at $\eta=1.3$ we must couple to higher order RSB. In principle, sideband cooling is not recoil-limited. However, there are several technical limiting factors which generates additional recoil heatings:

\begin{enumerate}[]
\item At the end of CRSC, we apply $2~\mu s$ optical pumping at  $s=0.8~(s=I/I_{sat})$ to clean up the population on $\ket{\uparrow}$ and $\ket{aux}$ states. One of the pumping beams is $120$ MHz red-detuned from the qubit Doppler transition. This step scatters 0.5 photons on average and heat up the ions: $\Delta{n}_{h}^{(1)} = 0.13~(\eta=0.78)$ and  $\Delta{n}_{h}^{(2)} =0.35~(\eta=1.3)$.

\item During the CRSC, the $120$ MHz red-detuned pumping beam off-resonantly couples to Doppler transition ($s=0.4$), corresponding to a 125 kHz photon scattering rate. This part of heating rates on the ions are $\dot{n}_{h1}^{(1)}=31$~ kHz~$(\eta=0.78)$ and $\dot{n}_{h1}^{(2)}=86$~kHz~$(\eta=1.3)$.

\item Off-resonant coupled to the carrier transition. The transition rate is $\Omega_{off} = \Omega_c^3/(\Omega_c^2+\Delta^2)$, and the photon scattering rate equals to $1.5\Omega_{off}/(2\pi)$. This part of heating on the ions are $\dot{n}_{h2}^{(1)}=3.0$~kHz~$(\eta=0.78)$ and $\dot{n}_{h2}^{(2)}=7.8$~kHz~$(\eta=1.3)$.

\end{enumerate}
In summary, extra recoil heating is the limiting factor for the final temperature. Adding a microwave assisted pumping from $\ket{aux}$ state to $\ket{\uparrow}$ state will highly suppress the off-resonant recoils. 



\subsection{Background gas collisions}
Finally, we consider the heating caused by background gas collisions. As our vacuum is mainly limited by the existence of hydrogen gases that are magnitudes more than other species, we assume that all collisions are induced by hydrogen. With room temperature (300K) hydrogen molecules colliding the beryllium ions under the hard-sphere collision model, we find that a mean energy of $E_{\mathrm{BGC}} = 6.2\times 10^{-22} J$. We further assume that $E_{\mathrm{BGC}}$ fills in equally to all M modes. As $M=3N$, for a single collision event, we have mean energy of:
\begin{equation}
\omega_{\mathrm{BGC}}=2\pi \times \frac{E_{\mathrm{BGC}}}{3N \times 2\pi\hbar}=2\pi \times 1.30\times 10^{10}~\mathrm{Hz}.
\end{equation}
This leads to $\bar{n}=47000$, which is far beyond the Doppler cooling limit. Such collisions usually lead to the melting of the entire chain when the Doppler cooling light is off, and it is observable: hydrogen molecules kick all the ions out, and no ion could survive in the trap. Therefore, the heating of background gas collisions is avoided by discarding the affected outliers, which are easily distinguishable and only account for a small portion of the total data. 

We measure the collision rate $R_{\mathrm{BGC}}$ in our vacuum system by counting the number of ion-ion hopping events per second. We find that on average $R_{\mathrm{BGC}}=0.07$ event/(ion$\times$ s), which is consistent with our estimation based on the vacuum pressure. So with 24 ions in the chain, we expect 0.168 collisions per second. Furthermore, as the time for a single run of an experiment is on the scale of 1 ms, we can assume a second is equivalent to 100 experiments. So we expect $1.6\times 10^{-3}$ collision per experiment, making the melting data account for $0.2\%$ of the total data. 




